# Entanglement measures and the Hilbert-Schmidt distance

Masanao Ozawa

*School of Informatics and Sciences, Nagoya University, Chikusa-ku, Nagoya 464-8601, Japan*

**Abstract**

In order to construct a measure of entanglement on the basis of a "distance" between two states, it is one of desirable properties that the "distance" is nonincreasing under every completely positive trace preserving map. Contrary to a recent claim, this letter shows that the Hilbert-Schmidt distance does not have this property.

As classical information arises from probability correlation between two random variables, quantum information arises from entanglement [1, 2]. Motivated by the finding of an entangled state which does not violate Bell's inequality, the problem of quantifying entanglement has received an increasing interest recently.

Vedral et. al. [3] proposed three necessary conditions that any measure of entanglement has to satisfy and showed that if a "distance" between two states has the property that it is nonincreasing under every completely positive trace preserving map (to be referred to as the CP nonexpansive property), the "distance" of a state to the set of disentangled states satisfies their conditions. It has been shown that the quantum relative entropy and the Bures metric have the CP nonexpansive property [3], and it has been conjectured that so does the Hilbert-Schmidt distance [4].

In the interesting Letter [5], Witte and Trucks claimed that the Hilbert-Schmidt distance really has the CP nonexpansive property and conjectured that the distance generates a measure of entanglement satisfying even the stronger condition posed later by Vedral and Plenio [4]. However, it can be readily seen that their suggested proof includes a serious gap. In this Letter, it will be shown that, contrary to their claim, the Hilbert-Schmidt distance does not have the CP nonexpansive property by presenting a counterexample.

Let $\mathcal{H} = \mathcal{H}_1 \otimes \mathcal{H}_2$ be the Hilbert space of a quantum system consisting of two subsystems with Hilbert spaces $\mathcal{H}_1$ and $\mathcal{H}_2$. We assume that $\mathcal{H}_1$ and $\mathcal{H}_2$ have the same finite dimension. We shall consider the notion of entanglement with respect to the above two subsystems. Let $\mathcal{T}$ be the set of density operators on $\mathcal{H}$. The set $\mathcal{D}$ of disentangled states is the set of all convex combinations of pure tensor product states. There are several requirements that every measure of entanglement, $E$, should satisfy [3, 4]:

(E1) $E(\sigma) = 0$ for all $\sigma \in \mathcal{D}$.

(E2) For any family of bounded operators $\{V_i\}$ of the form $V_i = A_i \otimes B_i$ such that $\sum_i V_i^\dagger V_i = I$,
 (a) $E(\sum_i V_i \sigma V_i^\dagger) \leq E(\sigma)$,
 (b) $\sum_i \text{Tr}[V_i \sigma_i V_i^\dagger] E(V_i \sigma_i V_i^\dagger / \text{Tr}[V_i \sigma_i V_i^\dagger]) \leq E(\sigma)$.

Condition (E1) ensures that disentangled states have a zero value of entanglement. Condition (E2) ensures that the amount of entanglement does not increase totally or in average by so-called purification procedures. Note that (E2-a) implies the following condition:

(E3) $E(\sigma) = E(U_1 \otimes U_2 \sigma U_1^\dagger \otimes U_2^\dagger)$ for all unitary operators $U_i$ on $\mathcal{H}_i$ for $i = 1, 2$.

Condition (E3) ensures that a local change of basis has no effect on the amount of entanglement.

Vedral et. al. [3] proposed the following general construction of the measure of entanglement $E$. Let $D : \mathcal{T} \times \mathcal{T} \to \mathbf{R}$ be a function satisfying the following conditions:

(D1) $D(\sigma, \rho) \geq 0$ and $D(\sigma, \sigma) = 0$ for any $\sigma, \rho \in \mathcal{T}$.

(D2) $D(\Theta \sigma, \Theta \rho) \leq D(\sigma, \rho)$ for any $\sigma, \rho \in \mathcal{T}$ and for any completely positive trace preserving map $\Theta$ on the space of operators on $\mathcal{H}$.

Condition (D1) ensures that $D$ has some properties of "distance". Condition (D2) ensures that the "distance" does not increase by any nonselective operations. Then, it is shown that the "distance" $E(\sigma)$ of a state $\sigma$ to the set $\mathcal{D}$ of disentangled states defined by

$$E(\sigma) = \inf_{\rho \in \mathcal{D}} D(\sigma, \rho) \tag{1}$$

satisfies conditions (E1) and (E2-a). It is shown that the quantum relative entropy and the Bures metric satisfy (D1) and (D2) [3], and it is conjectured that the Hilbert-Schmidt distance is a reasonable candidate of a "distance" to generate an entanglement measure [4]. Here, the Hilbert-Schmidt distance is defined by

$$D_{HS}(\sigma, \rho) = \|\sigma - \rho\|_{HS}^2 = \text{Tr}[(\sigma - \rho)^2]$$

for all $\sigma, \rho \in \mathcal{T}$, which satisfies (D1) since $\|\sigma - \rho\|_{HS}$ is a true metric.

Recently, Witte and Trucks [5] claimed that the Hilbert-Schmidt distance also satisfies (D2) and that the prospective measure of entanglement, $E_{HS}$, defined by

$$E_{HS}(\sigma) = \inf_{\rho \in \mathcal{D}} D_{HS}(\sigma, \rho)$$

satisfies (E1) and (E2-a).

It should be pointed out first that their suggested proof of condition (D2) for $D_{HS}$ is not justified. Let $f$ be a convex function on $(0, \infty)$ and let $f(0) = 0$. Let $\Phi$ be a trace preserving positive map on the space of operators such that $\|\Phi\| \leq 1$. Then, Lindblad's theorem [6] asserts that for every positive operator $A$ we have

$$\text{Tr}[f(\Phi A)] \leq \text{Tr}[f(A)], \tag{2}$$

where $f(A)$ is defined as usual through the spectral resolution of $A$. It is suggested that with the help of the above theorem it can be shown that

$$D_{HS}(\Theta \sigma, \Theta \rho) \leq D_{HS}(\sigma, \rho) \tag{3}$$

by regarding $D_{HS}$ as a convex function on $\mathcal{T}_+(\mathcal{H}) \oplus \mathcal{T}_+(\mathcal{H})$ for all positive mappings $\Theta$. However, it is not clear at all how $D_{HS}$ and $\Theta$ satisfy the assumptions of Lindblad's theorem.

Now, we shall show a counterexample to the claim that $D_{HS}$ satisfies condition (D2). Let $A$ and $B$ be $4 \times 4$ matrices defined by

$$A = \begin{pmatrix} 0 & 0 & 0 & 0 \\ 1 & 0 & 0 & 0 \\ 0 & 0 & 0 & 0 \\ 0 & 0 & 1 & 0 \end{pmatrix}, \quad B = \begin{pmatrix} 0 & 0 & 0 & 0 \\ 0 & 1 & 0 & 0 \\ 0 & 0 & 0 & 0 \\ 0 & 0 & 0 & 1 \end{pmatrix}.$$

Then we have

$$A^\dagger A = \begin{pmatrix} 1 & 0 & 0 & 0 \\ 0 & 0 & 0 & 0 \\ 0 & 0 & 1 & 0 \\ 0 & 0 & 0 & 0 \end{pmatrix}.$$

It follows that $A^\dagger A + B^\dagger B = I_4$ and hence

$$\Theta \sigma = A \sigma A^\dagger + B \sigma B^\dagger,$$

where $\sigma$ is arbitrary, defines a completely positive trace preserving map. Let $\sigma$ and $\rho$ be density matrices defined by

$$\sigma = \begin{pmatrix} 1/2 & 0 & 0 & 0 \\ 0 & 1/2 & 0 & 0 \\ 0 & 0 & 0 & 0 \\ 0 & 0 & 0 & 0 \end{pmatrix}, \quad \rho = \begin{pmatrix} 0 & 0 & 0 & 0 \\ 0 & 0 & 0 & 0 \\ 0 & 0 & 1/2 & 0 \\ 0 & 0 & 0 & 1/2 \end{pmatrix}.$$

Then we have

$$(\sigma - \rho)^2 = \begin{pmatrix} 1/4 & 0 & 0 & 0 \\ 0 & 1/4 & 0 & 0 \\ 0 & 0 & 1/4 & 0 \\ 0 & 0 & 0 & 1/4 \end{pmatrix}$$

and hence

$$D_{HS}(\sigma, \rho) = \text{Tr}[(\sigma - \rho)^2] = 1.$$

On the other hand, we have

$$A \sigma A^\dagger = \begin{pmatrix} 0 & 0 & 0 & 0 \\ 0 & 1/2 & 0 & 0 \\ 0 & 0 & 0 & 0 \\ 0 & 0 & 0 & 0 \end{pmatrix}, \quad B \sigma B^\dagger = \begin{pmatrix} 0 & 0 & 0 & 0 \\ 0 & 1/2 & 0 & 0 \\ 0 & 0 & 0 & 0 \\ 0 & 0 & 0 & 0 \end{pmatrix},$$

$$A \rho A^\dagger = \begin{pmatrix} 0 & 0 & 0 & 0 \\ 0 & 0 & 0 & 0 \\ 0 & 0 & 0 & 0 \\ 0 & 0 & 0 & 1/2 \end{pmatrix}, \quad B \rho B^\dagger = \begin{pmatrix} 0 & 0 & 0 & 0 \\ 0 & 0 & 0 & 0 \\ 0 & 0 & 0 & 0 \\ 0 & 0 & 0 & 1/2 \end{pmatrix}.$$

It follows that

$$(\Theta\sigma - \Theta\rho)^2 = \begin{pmatrix} 0 & 0 & 0 & 0 \\ 0 & 1 & 0 & 0 \\ 0 & 0 & 0 & 0 \\ 0 & 0 & 0 & 1 \end{pmatrix}$$

and hence

$$D_{HS}(\Theta\sigma, \Theta\rho) = \mathrm{Tr}[(\Theta\sigma - \Theta\rho)^2] = 2.$$

We conclude therefore

$$D_{HS}(\Theta\sigma, \Theta\rho) > D_{HS}(\sigma, \rho).$$

From the above counterexample, we conclude that the inequality

$$D_{HS}(\Theta\sigma, \Theta\rho) \leq D_{HS}(\sigma, \rho)$$

is not generally true for completely positive trace preserving maps $\Theta$. Therefore, it is still quite open whether $E_{HS}$ is a good candidate for an entanglement measure or not.

In order to obtain a tight bound for $D_{HS}(\Theta\sigma, \Theta\rho)$, we take advantage of Kadison's inequality [7]: If $\Phi$ is a positive map, then we have

$$\Phi(A)^2 \leq \|\Phi\| \, \Phi(A^2) \tag{4}$$

for all Hermitian $A$. Applying the above inequality to the positive trace preserving map $\Phi = \Theta$ and $A = \sigma - \rho$, we have

$$(\Theta\sigma - \Theta\rho)^2 \leq \|\Theta\| \, \Theta[(\sigma - \rho)^2].$$

By taking the trace of the both sides we obtain the following conclusion: *For any trace preserving positive map $\Theta$ and any states $\sigma$ and $\rho$, we have*

$$D_{HS}(\Theta\sigma, \Theta\rho) \leq \|\Theta\| D_{HS}(\sigma, \rho). \tag{5}$$

The previous example shows that the bound can be attained with $\|\Theta\| = 2$.

**Acknowledgements**

I thank V. Vedral and M. Murao for calling my attention to the present problem.